\documentclass[superscriptaddress,aps,showpacs,nofootinbib,twocolumn]{revtex4}

\usepackage{graphicx}
\usepackage{amsmath,amssymb}

\begin{document}

\title{Sneutrino brane inflation and leptogenesis}
\author{M. C. Bento}
\email{bento@sirius.ist.utl.pt}
\affiliation{Departamento de F\'{\i}sica and Centro de F\'{\i}sica das Interac\c{c}\~{o}es
Fundamentais, Instituto Superior T\'{e}cnico, Av. Rovisco Pais, 1049-001
Lisboa, Portugal}
\author{R. Gonz\'{a}lez Felipe}
\email{gonzalez@cfif.ist.utl.pt}
\affiliation{Departamento de F\'{\i}sica and Centro de F\'{\i}sica das Interac\c{c}\~{o}es
Fundamentais, Instituto Superior T\'{e}cnico, Av. Rovisco Pais, 1049-001
Lisboa, Portugal}
\author{N. M. C. Santos}
\email{ncsantos@cfif.ist.utl.pt}
\affiliation{Departamento de F\'{\i}sica and Centro de F\'{\i}sica das Interac\c{c}\~{o}es
Fundamentais, Instituto Superior T\'{e}cnico, Av. Rovisco Pais, 1049-001
Lisboa, Portugal}

\begin{abstract}
Modifications to the Friedmann equation in brane cosmology can have important
implications for early universe phenomena such as inflation and the generation
of the baryon asymmetry. We study a simple scenario of chaotic brane inflation
where, in a minimal supersymmetric seesaw model, the scalar superpartner of a
heavy singlet Majorana neutrino drives inflation and, simultaneously, generates
the required lepton asymmetry through its direct out-of-equilibrium decays
after the inflationary era. For a gravitino mass in the range $m_{3/2} \simeq$
100~GeV~-~1~TeV, we find that successful nucleosynthesis and leptogenesis in
this framework require that the 5D Planck mass is in the range $M_{5} \simeq
10^{10}-10^{13}$~GeV and the reheating temperature $T_{rh} \simeq
10^{6}-10^{8}$~GeV.
\end{abstract}


\maketitle

\section{Introduction}
\label{one}

Today there is a wide consensus that the early universe underwent a period of
cosmological inflation \cite{Lyth:1998xn}. Inflationary era can be regarded as
a necessary stage, responsible not only for the observed flatness, homogeneity
and isotropy of the present universe, but also for the origin of the density
fluctuations as observed by the Cosmic Background Explorer (COBE) and, more
recently, the Wilkinson Microwave Anisotropy Probe (WMAP)
satellites~\cite{Bennett:2003bz}. At the end of inflation, the universe was in
a cold and low-entropy state and it must has been subsequently reheated to
become a high-entropy and radiation-dominated universe. Such a reheating
process could occur, for instance, through the coherent oscillations of the
inflaton field about the minimum of the potential until the age of the universe
equals the lifetime of the inflaton. The latter decays into ordinary particles,
which then scatter and thermalize. Besides entropy creation, the right
abundance of baryons must be created after the inflationary epoch. This usually
poses serious problems in constructing particle physics models which lead
simultaneously to a successful inflationary and baryogenesis scenario. In
particular, the reheating temperature is typically too low when compared with
the grand unification scale, at which baryogenesis is expected to take place in
the simplest GUTs. Moreover, any preexisting baryon asymmetry would be erased
by the anomalous sphaleron processes \cite{Kuzmin:1985mm} unless an initial
$B-L$ asymmetry is generated.

Another major obstacle in constructing viable supergravity-inspired
cosmological models is the overproduction of gravitinos. In conventional
scenarios, the gravitino mass is expected to be comparable to the masses of the
supersymmetric partners of the standard model particles and, therefore,
$m_{3/2}\lesssim$ a few TeV in order to solve the gauge hierarchy problem.
Since the gravitino coupling to matter is suppressed by the Planck mass $M_{P}$
, its lifetime is $\tau_{3/2} \sim M_{P}^{2}/m_{3/2}^{3} \sim 10^{8}(100$~GeV
$/m_{3/2})^{3}$~s. During the reheating phase gravitinos can be thermally
produced through scatterings in the plasma. However, if they are overproduced
after inflation, their decay products could put at risk the successful
predictions of primordial nucleosynthesis~\cite{Khlopov:pf,Cyburt:2002uv}.
Since in standard cosmology their abundance is proportional to the reheating
temperature, $T_{rh}\,$, constraints from big bang nucleosynthesis (BBN) yield
a stringent upper bound on the allowed $T_{rh}$ after inflation: $T_{rh}
\lesssim 10^{7}-10^{10}$~GeV for 100~GeV $\lesssim m_{3/2} \lesssim 1$~TeV
\cite{Cyburt:2002uv}.

Among the current chaotic inflationary scenarios in supersymmetric seesaw
theories~\cite{Murayama:1992ua,Hamaguchi:2001gw,Ellis:2003sq}, inflation driven
by the scalar superpartner of the right-handed Majorana neutrino is one of the
simplest and most economical ones. In this context, the heavy singlet neutrinos
are naturally invoked to give masses to the light neutrinos through the seesaw
mechanism~\cite{seesaw}. Moreover, their superpartners - the sneutrino fields -
can play the role of the inflaton. Also, if $CP$ is violated the sneutrino
decays will create a lepton asymmetry, which is then converted into a baryon
asymmetry by the electroweak sphalerons. This is indeed an appealing scenario,
since cosmology and particle physics merge together to make predictions about
the early universe and the low-energy physics that we test today.

There is, however, a drawback in the above-mentioned framework. In the usual
chaotic inflation scenario based on standard cosmology, super-Planckian
inflaton field values $\sim 3 M_{P}$ are typically required to allow for a
sufficiently long period of inflation (the so-called $\eta$ problem). Thus one
expects nonrenormalizable quantum corrections of the order of
$\mathcal{O}[(\phi /M_{P})^{n}]$ (with $n > 4$) to destroy the flatness of the
potential necessary for successful inflation. A possible way out of this
situation is to consider, for instance, higher-dimensional cosmological models,
where our four-dimensional world is viewed as a 3-brane embedded in a
higher-dimensional bulk.

A remarkable feature of brane cosmology is the modification of the expansion
rate of the universe $H$ before the nucleosynthesis era \cite{Binetruy:1999ut}.
While in standard cosmology the expansion rate scales with the energy density
$\rho$ as $H\propto\sqrt{\rho}$, this dependence becomes $H\propto\rho$ at very
high energies in brane cosmology. This behavior, which appears to be quite
generic and not specific to Randall-Sundrum braneworld
scenarios~\cite{Randall:1999vf}, may have drastic consequences on early
universe phenomena such as inflation and the generation of the baryon
asymmetry. In particular, modifications to the Friedmann equation not only ease
the conditions for slow-roll inflation but also enable the simplest chaotic
inflation models to inflate at field values far below $M_{P},$ thus avoiding
well-known difficulties with higher-order nonrenormalizable terms. Another
important difference between standard and brane cosmologies is in the
predictions for gravitino production. For a given value of the brane tension
or, equivalently, of the 5D Planck mass, $M_{5}$, the gravitino abundance in
the brane decreases as $T_{rh}$ increases. Therefore, in contrast to standard
cosmology, BBN constraints in the brane scenario imply a lower (rather than an
upper) bound on the reheating temperature.

The aim of this paper is combines the above ideas, i.e. chaotic inflation and
direct leptogenesis through sneutrino decays in the braneworld
context~\cite{Dvali:1999gf}. More precisely, we study a simple scenario of
chaotic brane inflation where, in a minimal supersymmetric seesaw model, the
scalar superpartner of a heavy singlet Majorana neutrino drives inflation and,
simultaneously, generates the required lepton asymmetry through its direct
out-of-equilibrium decays after the inflationary era. This requires the
reheating temperature $T_{rh}$ to be smaller than the sneutrino inflaton mass,
$M_1\,$. In this framework, there exists a direct connection between the brane
inflationary era, the reheating of the universe, leptogenesis from sneutrino
decays and the light neutrino properties, which allows us to strongly constrain
the fundamental 5D Planck mass and the reheating temperature of the universe.

We shall not consider here the case where leptogenesis is purely thermal, i.e.
when $T_{rh} > M_1$. In this case, any lepton asymmetry generated through the
sneutrino inflaton decays is erased by thermal effects, and therefore,
leptogenesis is driven by the out-of-equilibrium decays of the heavy singlet
Majorana neutrinos and sneutrinos thermally created.

\section{Sneutrino braneworld inflation}

We consider the simplest scenario where heavy right-handed neutrinos
$N_i\,,\,i=1,2,3\,$, with masses $M_i\,$, are added to the usual particle
content of the minimal supersymmetric standard model. In what follows we assume
that the seesaw mechanism~\cite{seesaw} is operative in the brane scenario and
gives masses to the light neutrinos. We remark however that in the presence of
extra dimensions other mechanisms are viable as well. In particular, it is
possible to generate light neutrino masses without the need for a superheavy
mass scale~\cite{Dienes:1998sb,Arkani-Hamed:1998vp}. We also assume that the
lightest right-handed sneutrino $\widetilde{N}_1$ acts as an inflaton with a
potential simply given by the mass term
\begin{align}
    V=\frac{1}{2} M_1^2 \widetilde{N}_1^2\,. \label{pot}
\end{align}
For simplicity we neglect the dynamics of the heavier sneutrinos
$\widetilde{N}_{2,3}\,$. Under these assumptions, the model reduces to the
simplest braneworld chaotic inflation scenario with a quadratic potential of
the form $V =\frac{1}{2} m^2 \phi^2$.

We start by reviewing the constraints resulting from WMAP bounds on
inflationary observables. In standard cosmology, the requirement that
perturbations have the observed amplitude fixes the inflaton mass: $m \simeq
10^{13}$~GeV. However, due to the presence of the brane tension, $\lambda$, the
Friedmann equation in the braneworld context receives an additional term
quadratic in the density~\cite{Binetruy:1999ut},
\begin{align}
\label{Fried} H^2 = \frac{8\pi}{3 M_P^2} ~ \rho ~ \left(1 +
\frac{\rho}{2\lambda} \right) ~,
\end{align}
where $M_P=1.22 \times 10^{19}$~GeV is the 4D Planck mass. In the above
expression for $H$, we have set the four-dimensional cosmological constant to
zero and assumed that inflation rapidly makes any dark radiation term
negligible. Notice that Eq.~(\ref{Fried}) reduces to the usual Friedmann
equation at sufficiently low energies, \mbox{$\rho \ll \lambda$}. The brane
tension relates the Planck mass in four and five  dimensions via
\begin{align}
M_P = \sqrt{\frac{3}{4 \pi}} \frac{M_5^3}{\sqrt{\lambda}}~. \label{eq:MP}
\end{align}

Successful big bang nucleosynthesis requires that the change in the expansion
rate due to the new terms in the Friedmann equation be sufficiently small at
scales $\sim \mathcal{O}$(MeV). This implies the lower bound $ \lambda \gtrsim
(1~\mbox{MeV})^4$, or using Eq.~(\ref{eq:MP}), $M_5 \gtrsim
40~\mbox{TeV}$~\cite{Cline:1999ts}. A more stringent bound, $M_5 \gtrsim
10^5~\mbox{TeV}~, $ can be obtained by requiring the theory to reduce to
Newtonian gravity on scales larger than 1 mm~\cite{Maartens:1999hf}.

In the slow-roll approximation, the total number  of $e$-folds during inflation
is given by~\cite{Maartens:1999hf}
\begin{align}
\label{efolds} N \simeq - \frac{8\pi}{M_P^2} \int^{\phi_{f}}_{\phi_{i}}
\frac{V}{V'} \left( 1+\frac{V}{2\lambda} \right) d\phi ~,
\end{align}
where $\phi_{i}$ and $\phi_{f}$ are the values of the scalar field at the
beginning and at the end of the expansion, respectively. The value $\phi_f$ can
be computed from the condition ${\rm max}\{\epsilon(\phi_f),|\eta(\phi_f)|\}=
1$, where $\epsilon$ and $\eta$ are the slow-roll parameters, given by
\begin{align}
\epsilon & \equiv  \frac{M_P^2}{16\pi} ~ \left(
    \frac{V'}{V} \right)^2 \; \frac{1 + V/\lambda}{\left(1 +
    V/2\lambda \right)^2} ~, \\
\eta & \equiv  \frac{M_P^2}{8\pi} ~ \frac{V''}{V} \;
    \frac{1}{1+V/2\lambda}  ~.
\end{align}
Other key parameters during inflation are the spectra of scalar
\cite{Maartens:1999hf} and tensor \cite{Langlois:2000ns} perturbations at the
Hubble radius crossing:
\begin{align}
\label{scalamp} A_s^2 & =   \frac{512 \pi}{75 M_P^6} ~ \frac{V^3}{V'^2} ~
\left( 1 +
\frac{V}{2\lambda} \right)^3 ~, \\
A_t^2 & =  \frac{32}{75 M_P^4} ~ V \left( 1 + \frac{V}{2\lambda} \right)~F^2 ~.
\label{tensamp}
\end{align}
Here
\begin{align}
F^2(x) &= \left[\sqrt{1+x^2} - x^2 \sinh^{-1} \frac{1}{x} \right]^{-1}~,\\
x &=\left[ \frac{2 V}{\lambda}\left(1 + \frac{V}{2 \lambda}\right)
\right]^{1/2} ~. \label{functionF}
\end{align}
In the low-energy limit, when $V\ll \lambda$ (i.e. $x \ll 1$), one has $F^2
\approx 1$, whereas in the high-energy limit, $V\gg \lambda$, we obtain $F^2
\approx 3 V/2\lambda\,$.

The scale dependence of the scalar perturbations is described by the spectral
tilt
\begin{align}
n_s-1 \equiv \frac{d \ln A_{s}^2}{d \ln k}\simeq -6 \epsilon + 2 \eta~,
\end{align}
and its running is given by
\begin{align}
\alpha_s \equiv \frac{d n_s}{d \ln k}\simeq 16 \epsilon \eta -18 \epsilon^2 -
2\xi~,
\end{align}
where
\begin{align}
\xi \equiv \frac{M_{P}^4}{(8\pi)^2}\frac{V' V'''}{V^2} \frac{1}{\left(1+{V/ 2
\lambda}\right)^2}~. \label{eq:xi}
\end{align}
Finally, the tensor power spectrum amplitude can be parametrized by the
tensor-to-scalar ratio
\begin{align}
r_s \equiv 16 ~ \frac{A_{t}^2}{A_{s}^2} ~,
\end{align}
which is consistent with the normalization of Ref.~\cite{Peiris:2003ff} in the
low-energy limit.

For the simple quadratic potential of Eq.~(\ref{pot}) one gets in the
high-energy approximation\footnote{Our expressions do not coincide with the
ones of Ref.~\cite{Liddle:2003gw} because we take the condition $\epsilon=1$
rather than $\eta=1$ to find $\phi_f$. The difference in the numerical values
is however insignificant.}
\begin{align}
n_s&\simeq \frac{2(N_\star -2)}{2 N_\star+1}~,\\
\alpha_s  &\simeq  -\frac{10}{(2 N_\star +1 )^2}~,\\
r_s &\simeq \frac{24}{2 N_\star+1}~, \label{nrmod}
\end{align}
where $N_\star$ is the number of $e$-folds before the end of inflation at which
observable perturbations are generated\footnote{In standard cosmology, $N_\star
= 55$ is found to be a reasonable fiducial value. This number is expected to be
higher in the brane scenario~\cite{Dodelson:2003vq}; in Ref.~\cite{Wang:2003qr}
the upper bound $N_\star<75$ is found.}. Hence, taking e.g. $N_\star=60$, we
get $n_s\approx 0.96,~\alpha_s\approx -6.83 \times 10^{-4},~ r_s\approx 0.2 $,
well within the WMAP bounds on these quantities~\cite{Peiris:2003ff}
\begin{align}
0.94 \leq n_s \leq 1.01~,~-0.02 \leq \alpha_s \leq 0.02~,~r_s \leq 0.35~,
\label{eq:WMAPConst2}
\end{align}
at $95\%$ CL.

Using Eq.~(\ref{scalamp}) the mass parameter $m$ can be expressed in terms of
$A_s$ and $N_\star$ as
\begin{align}
\label{m} m=\left(\frac{75\sqrt{3}\pi A_s^2}{8\sqrt{2} (1+2
N_\star)^{5/2}}\right)^{1/3} M_5~.
\end{align}
Inserting into this expression the COBE normalization $A_s(\phi_\star)\approx 2
\times 10^{-5}$ and using $N_\star=60$, we obtain
\begin{align}
\label{mnum}
m\approx 4.5 \times 10^{-5} ~M_5~.
\end{align}
Let us note that the above analysis has been done assuming that the high energy
approximation is valid, i.e. $ V/\lambda \gg 1$. It is easy to show that this
requirement imposes a rather weak constraint on the fundamental scale $M_5$,
namely, $M_5 \ll 10^{17}~\mbox{GeV}$. It is also worth remarking that, in
contrast to the standard cosmology case, it is possible to obtain sufficient
inflation in the braneworld context for sub-Planckian initial values of the
inflaton field. Indeed, taking for instance $N=70$, the minimum number of
$e$-folds required to solve the initial conditions problems of standard
cosmology, one estimates $\phi_i \approx 3 \times 10^2~M_5$, which when
combined with the above bound on $M_5$ implies $\phi_i < M_P$.

\section{Reheating and direct leptogenesis from sneutrino decays}

After inflation, the cold inflaton-dominated universe undergoes a phase of
reheating, during which the inflaton decays into normal particles and the
universe becomes radiation-dominated. Of particular interest is the reheating
temperature, $T_{rh}$, which is defined by assuming an instantaneous conversion
of the inflaton energy into radiation, when the decay width of the inflaton,
$\Gamma_\phi$, equals the expansion rate of the universe, $H$. In the
braneworld scenario, $H$ is given by
\begin{align}
\label{Hb} H \simeq \sqrt{\frac{4\pi}{3 \lambda}}\frac{\rho}{M_P}=
\frac{4\pi}{3}\frac{\rho}{M_5^3}~,
\end{align}
in the high energy approximation. If the total inflaton energy density is
instantaneously converted into radiation, then we can identify
$\rho=\rho_R=(\pi^2/30)\,g_\ast\, T^4\,$, where $g_\ast$ is the effective
number of relativistic degrees of freedom at the temperature $T$. In
particular, $g_\ast = 915/4 = 228.75$ in the MSSM for temperatures above the
SUSY breaking scale $\sim \mathcal{O}$(TeV). Thus we obtain
\begin{align}
H\simeq \frac{2 \pi^3 g_\ast}{45} \frac{T^4}{M_5^3}~. \label{Hreh}
\end{align}
The condition $H(T_{rh})=\Gamma_\phi$ leads then to the relation
\begin{align}
T_{rh}= \left(\frac{45}{2 \pi^3 g_\ast} \Gamma_\phi M_5^3\right)^{1/4}~.
\label{trh}
\end{align}
In the sneutrino inflation scenario we are considering, $\phi\equiv \tilde
N_1\,,~m \equiv M_1$ and
\begin{align}
\Gamma_\phi \equiv \Gamma_{N_1}=\frac{1}{4\pi}\, (Y_\nu Y^\dagger_\nu)_{11}
M_1\,, \label{gamma}
\end{align}
where $Y_\nu$ is the Dirac neutrino Yukawa matrix. Substituting
Eq.~(\ref{gamma}) into (\ref{trh}), we get
\begin{align}
T_{rh}=\left[\frac{45}{8\pi^4 g_\ast} (Y_\nu Y^\dagger_\nu)_{11} M_1
M_5^3\right]^{1/4}~. \label{trh1}
\end{align}

At the end of inflation, when the Hubble parameter becomes smaller than $M_1$,
the inflaton field ${\widetilde N}_1$ begins to oscillate coherently around the
minimum of the potential. If $CP$ is not conserved, the decays of ${\widetilde
N}_1$ into leptons, Higgs and the corresponding antiparticles can produce a net
lepton asymmetry. We require such decays to occur out of equilibrium, i.e.
$T_{rh} < M_1$, so that leptogenesis is driven by the decays of cold sneutrino
inflatons and the produced lepton asymmetry is not washed out by lepton-number
violating interactions mediated by $N_1$. The above requirement, combined with
Eqs.~(\ref{trh1}) and (\ref{mnum}), leads to the following constraint on the
Dirac neutrino Yukawa couplings
\begin{align}
(Y_\nu Y^\dagger_\nu)_{11} \lesssim 3.5 \times 10^{-10}~,
\end{align}
which is independent of the fundamental scale $M_5$. Therefore, as in the
standard sneutrino inflationary scenario~\cite{Ellis:2003sq}, the quantity
$(Y_\nu Y^\dagger_\nu)_{11}$ is required to be very small.

The lepton number density generated by the sneutrino condensate is given by
\begin{align}
    n_L = \epsilon_1 \frac{\rho}{M_1}~,
\end{align}
where the parameter $\epsilon_1$ denotes the $CP$ asymmetry in the ${\widetilde
N}_1$ decays. Recalling that the entropy density is $s = (2 \pi^2 g_*/45)\,T^3
\,$, the lepton-to-entropy ratio can be written
as~\cite{Murayama:1992ua,Hamaguchi:2001gw}
\begin{align}
Y_L\equiv \frac{n_L}{s}= \frac{3}{4}\, \epsilon_1 \frac{T_{rh}}{M_1}~,
\label{yl}
\end{align}
with $T_{rh}$ given by Eq.~(\ref{trh1}).

Assuming the mass hierarchy $M_1 \ll M_2\,, M_3\,$ for the heavy Majorana
neutrinos, one has~\cite{Hamaguchi:2001gw}
\begin{align}
\epsilon_1 \simeq \frac{3}{8 \pi} \frac{M_1}{v^2 \sin^2 \beta}
\frac{\mbox{Im}\, [Y_\nu \mathcal{M}_\nu^\ast Y_\nu^T]_{11}}{(Y_\nu
Y^\dagger_\nu)_{11}}\,,
\end{align}
where $\mathcal{M}_\nu$ is the light neutrino effective mass matrix, $v \simeq
174$~GeV and $\tan \beta$ is the ratio of the vacuum expectation values of the
two Higgs doublets of the MSSM. It is convenient to parametrize the above
expression in the form
\begin{align}
\epsilon_1=\epsilon_1^{\rm max} \sin \delta_L~, \label{eps1}
\end{align}
where $\delta_L$ is an effective leptogenesis phase and $\epsilon_1^{\rm max}$
is the maximal asymmetry. In particular, for hierarchical light neutrinos with
$m_1 \simeq 0 \ll m_2 \simeq \sqrt{\Delta m_{\rm sol}^2} \ll m_3 \simeq
\sqrt{\Delta m_{\rm atm}^2}$, one obtains\footnote{For a more accurate bound
and a detailed discussion see e.g.
Ref.~\cite{Hambye:2003rt}.}~\cite{Davidson:2002qv}
\begin{align}
\epsilon_1^{\rm max} = \frac{3}{8\pi} \frac{M_1\sqrt{\Delta m_{\rm
atm}^2}}{v^2\sin^2\beta}~,
\end{align}
where the squared mass differences, measured in solar and atmospheric neutrino
oscillation experiments, are $\Delta m_{\rm sol}^2 \simeq 7.1\times
10^{-5}~\mbox{eV}^2 $ and $\Delta m_{\rm atm}^2 \simeq 2.6\times
10^{-3}~\mbox{eV}^2 $. With $\sin \beta \simeq 1$ (large $\tan \beta$ regime)
the maximal $CP$ asymmetry is then given by
\begin{align}
\epsilon_1^{\rm max} \simeq 2\times 10^{-10} \left(\frac{M_1}{10^6\,
\mbox{GeV}}\right)\,. \label{eps1max}
\end{align}

The lepton asymmetry produced before the electroweak phase transition is
partially converted into a baryon asymmetry via the sphaleron effects,
\cite{Kuzmin:1985mm}
\begin{align}
Y_B \equiv \frac{n_B}{\rm s}= \xi Y_L~, \label{yb}
\end{align}
where $\xi=-8/23$ for the MSSM~\cite{Khlebnikov:sr}.

From Eqs.~(\ref{yl}), (\ref{eps1}), (\ref{eps1max}) and (\ref{yb}), we obtain
\begin{align}
Y_B = 5.3 \times 10^{-11} \sin \delta_L \left(\frac{T_{rh}}{10^6\,
\mbox{GeV}}\right)\,. \label{yb2}
\end{align}

From the observational side, WMAP bounds on the baryon-to-photon ratio, $\eta_B
\equiv n_B/n_\gamma\,$, imply~\cite{Bennett:2003bz}
\begin{align}
\eta_B = 6.1^{+0.3}_{-0.2} \times 10^{-10}~.
\label{nbwmap}
\end{align}
Recalling that $s \simeq 7.04\, n_\gamma$, Eqs.~(\ref{yb2}) and (\ref{nbwmap})
lead to
\begin{align}
T_{rh} = \frac{1.6 \times 10^6}{|\sin \delta_L|}~\mbox{GeV}~, \label{ntrhb}
\end{align}
which yields the following lower bound on the reheating temperature
\begin{align}
T_{rh} \gtrsim 1.6 \times 10^6~\mbox{GeV}~, \label{trhlow}
\end{align}
for $\epsilon_1=\epsilon_1^{\rm max}$.

\section{Constraints on gravitino production}

As discussed in the Introduction, a viable cosmological supergravity-inspired
scenario has to avoid the so-called gravitino problem, i.e. for unstable
gravitinos, their decay products should not alter the BBN predictions for the
abundance of light elements in the universe. This requirement is usually
translated into an upper bound on the gravitino abundance~\cite{Cyburt:2002uv}:
\begin{align}
\eta_{3/2} \equiv \frac{n_{3/2}}{n_\gamma} \lesssim \frac{\zeta_{\rm
max}}{m_{3/2}}~, \label{bga}
\end{align}
where $m_{3/2}$ is the gravitino mass, $n_\gamma(T)=2\zeta(3) T^3/\pi^2$ is the
photon density and $\zeta_{\rm max}$ is a parameter that accounts for the
maximum gravitino abundance allowed by the BBN predictions. In particular, for
$m_{3/2}\simeq 100~\mbox{GeV}$ and a gravitino lifetime $\tau_{3/2}\simeq 10^8$
s, one has~\cite{Cyburt:2002uv}
\begin{align}
\zeta_{\rm max} \simeq 5\times 10^{-12}~\mbox{GeV}~, \label{100gev}
\end{align}
while for $m_{3/2}\simeq 1~\mbox{TeV}$ and $\tau_{3/2}\simeq 10^5$~s,
\begin{align}
\zeta_{\rm max} \simeq 10^{-8}~\mbox{GeV}~. \label{1tev}
\end{align}

The gravitino abundance at a given temperature $T < T_{rh}$ can be obtained
from the Boltzmann equation
\begin{align}
\frac{d n_{3/2}}{dt} + 3 H n_{3/2} = C_{3/2} (T)~,
\label{bolt}
\end{align}
where $C_{3/2}(T)$ is the collision term. Assuming constant entropy, the
integration of Eq.~(\ref{bolt}) yields
\begin{align}
\eta_{3/2}(T)=\frac{g_\ast(T)}{g_\ast(T_{rh})} \frac{C_{3/2}(T_{rh})}{H(T_{rh})
n_\gamma(T_{rh})}~, \label{gabft}
\end{align}
where $g_\ast(T)=43/11 \simeq 3.91$ for $T<1~\mbox{MeV}$, whereas
$g_\ast(T_{rh})=228.75\,$. The collision term is given by \cite{Bolz:2000fu}
\begin{align}
C_{3/2}(T)\simeq \alpha (T)\left( 1+ \beta (T)
\frac{m_{{\tilde g}^2}}{m_{3/2}^2}\right)
\frac{T^6}{M_P^2}~,
\label{c32}
\end{align}
where $m_{{\tilde g}}$ is the low-energy gluino mass; $\alpha (T)$ and $\beta
(T)$ are slowly-varying functions of the temperature. For $T_{rh}\sim
10^6~\mbox{GeV}$ (cf. Eq.~(\ref{trhlow})) we estimate $\alpha (T_{rh}) \simeq
2.38$ and $\beta (T_{rh}) \simeq 0.13$.

Substituting Eqs.~(\ref{Hreh}) and (\ref{c32}) into Eq.~(\ref{gabft}), we
obtain for the gravitino abundance
\begin{align}
\eta_{3/2}\simeq 5.3 \times 10^{-4} \left( 1+0.13
 \frac{m_{{\tilde g}^2}}{m_{3/2}^2} \right) \frac{M_5^3}{T_{rh}~M_P^2}~.
\label{gabm5}
\end{align}

Taking into account the BBN gravitino abundance bound (see Eqs.
(\ref{bga})-(\ref{1tev})) we find
\begin{eqnarray}
T_{rh}& \gtrsim & 970 \left( \frac{M_5}{10^{10}~\mbox{GeV}}\right)^3\quad
\mbox{for}~ m_{3/2}=100~\mbox{GeV}~,\nonumber\\
T_{rh}& \gtrsim & 0.4 \left( \frac{M_5}{10^{10}~\mbox{GeV}}\right)^3\quad
\mbox{for}~ m_{3/2}=1~\mbox{TeV}~, \label{b1t}
\end{eqnarray}
for a typical gluino mass of $m_{\tilde g}=1~\mbox{TeV}$. We notice that in
contrast with the standard cosmology case, where the gravitino production
imposes an upper bound on the reheating temperature of the universe, the BBN
constraints in the braneworld scenario imply a lower bound on $T_{rh}$ instead.

Combining the requirement to avoid thermal leptogenesis, $T_{rh}< M_1$, with
Eqs.~(\ref{mnum}) and (\ref{b1t}), we get
\begin{eqnarray}
M_5 &\lesssim & 2.1 \times 10^{11} ~\mbox{GeV}\quad \mbox{for}
~ m_{3/2}=100~\mbox{GeV}~,\nonumber\\
M_5 &\lesssim & 1.1 \times 10^{13} ~\mbox{GeV}\quad \mbox{for} ~
m_{3/2}=1~\mbox{TeV}~. \label{m5b}
\end{eqnarray}
We note that these bounds are more stringent than the one imposed by the
validity of the high energy approximation ($M_5 \lesssim 10^{17}$~GeV). The
above limits can also be translated into upper bounds on the lightest
right-handed Majorana neutrino mass:
\begin{eqnarray}
M_1 &\lesssim & 9.6\times 10^6 ~\mbox{GeV}\quad   \mbox{for}
~ m_{3/2}=100~\mbox{GeV}~,\nonumber\\
M_1 &\lesssim & 4.7 \times 10^{8} ~\mbox{GeV}\quad \mbox{for} ~
m_{3/2}=1~\mbox{TeV}~.
\end{eqnarray}

\begin{figure*}[t]
\includegraphics[width=8.5cm]{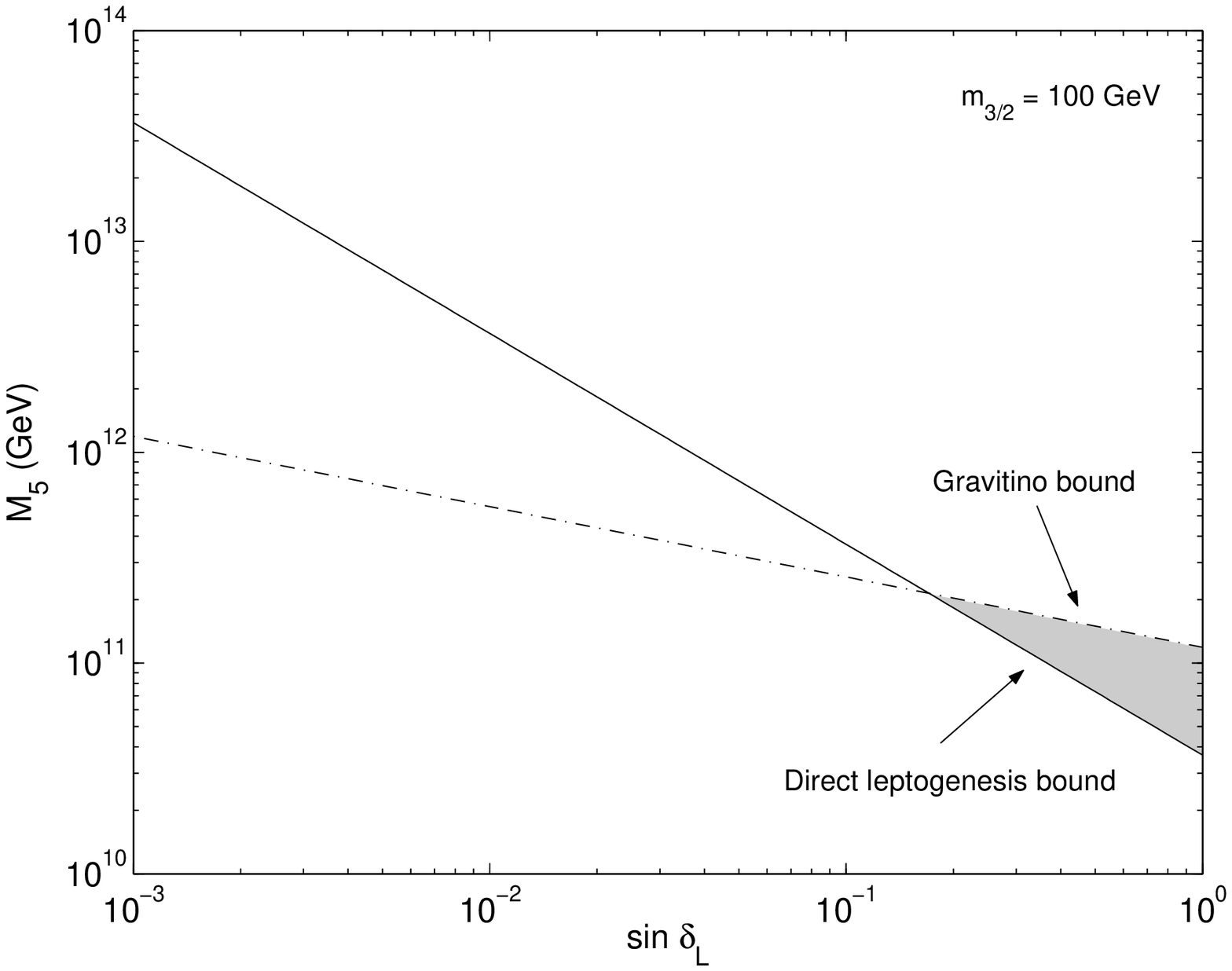}
\includegraphics[width=8.5cm]{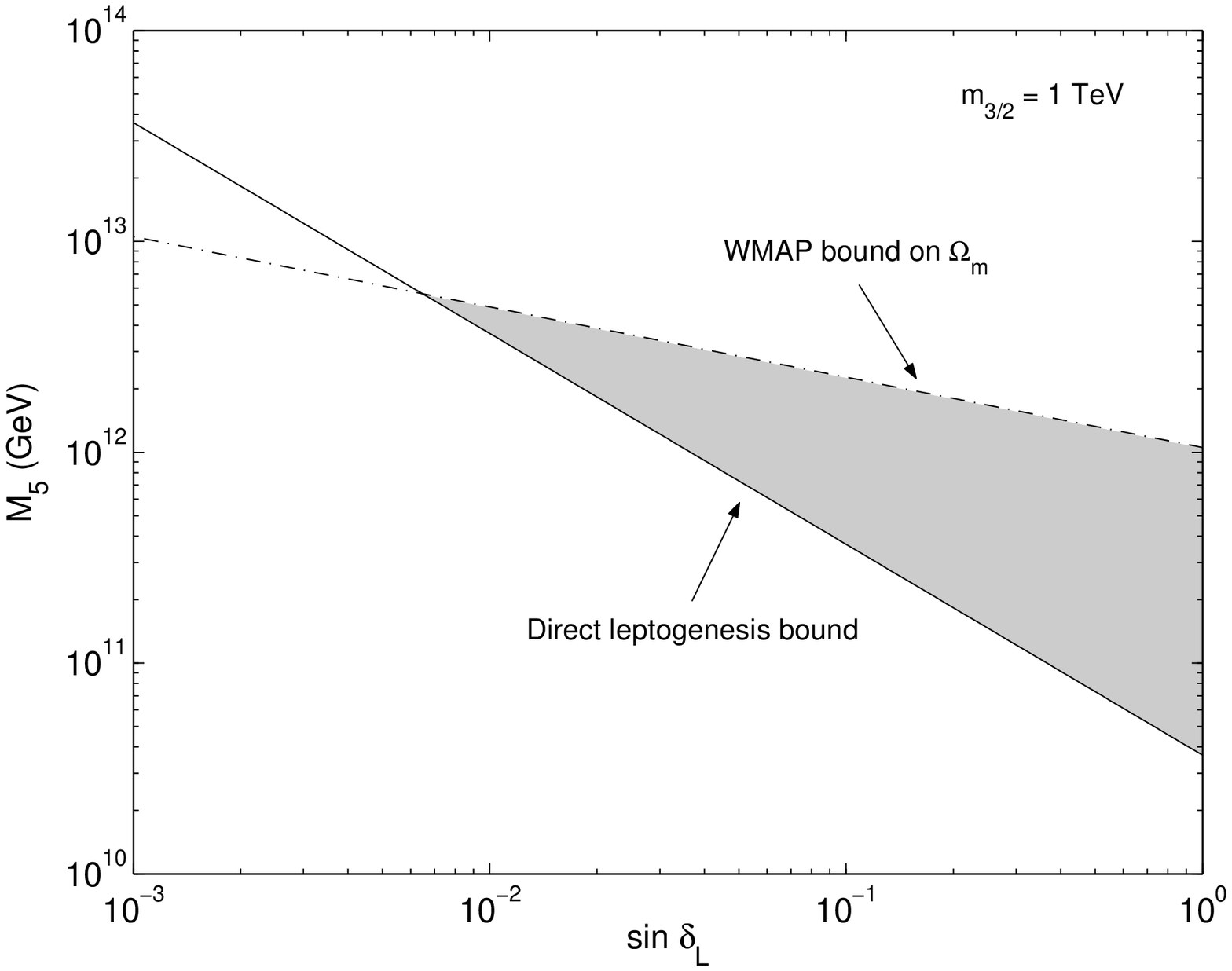}
\caption{The 5D Planck mass $M_5$ as a function of the effective leptogenesis
phase $\sin \delta_L$ for $m_{3/2}=100~\mbox{GeV}$ and $m_{3/2}=1~\mbox{TeV}$.
The shaded area corresponds to the allowed region, taking into account the BBN
constraints on gravitino production, the WMAP bounds on $\eta_B$ and
$\Omega_m$, and assuming direct leptogenesis from sneutrino decays.}
\label{fig1}
\end{figure*}
\begin{figure*}[t]
\includegraphics[width=8.5cm]{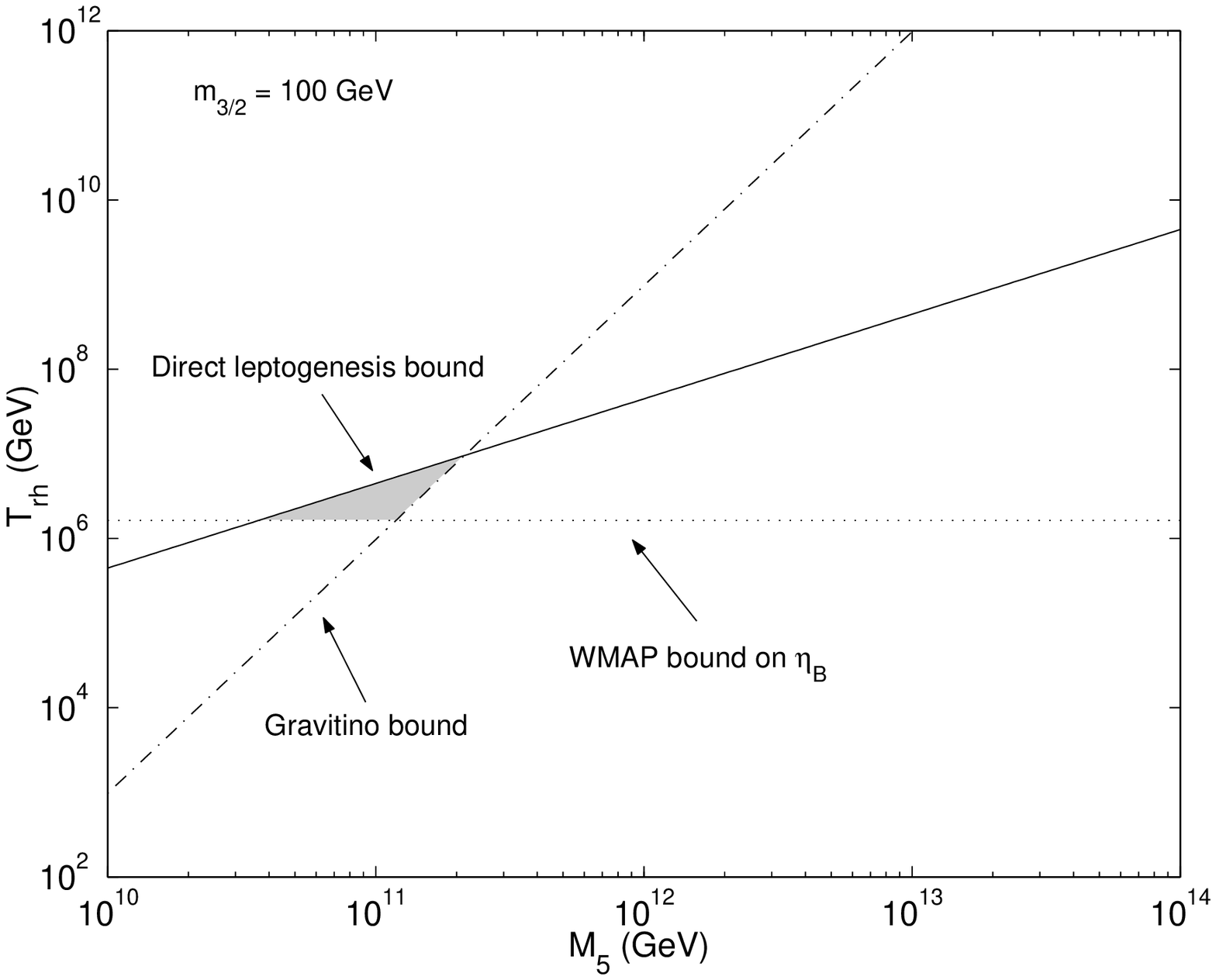}
\includegraphics[width=8.5cm]{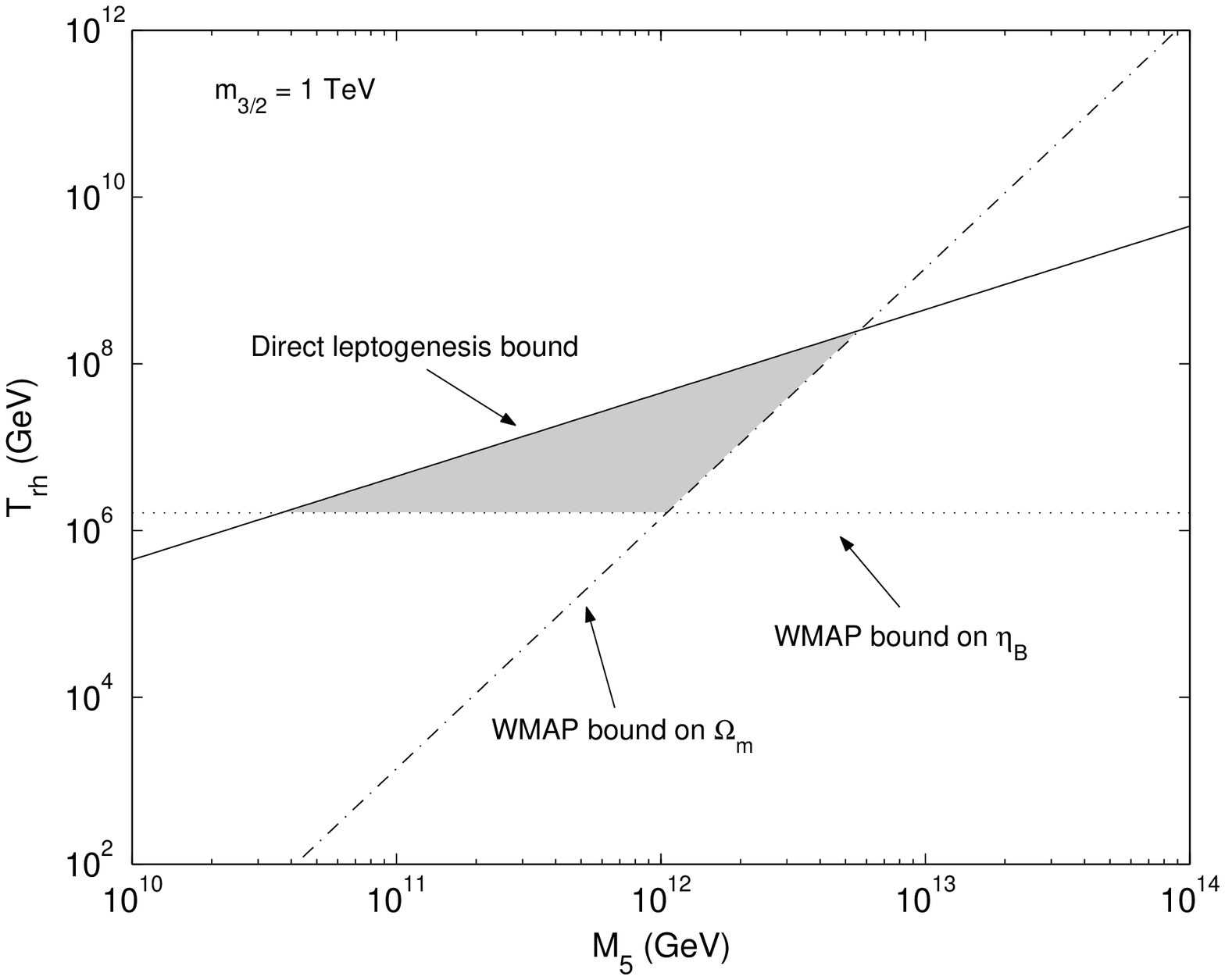}
\caption{The reheating temperature $T_{rh}$  as a function of $M_5$ for
$m_{3/2}=100~\mbox{GeV}$ and $m_{3/2}=1~\mbox{TeV}$ as derived from the BBN
gravitino constraints, direct leptogenesis and the WMAP bounds on $\eta_B$ and
$\Omega_m$. The shaded area corresponds to the region compatible with the above
bounds.} \label{fig2}
\end{figure*}

One should check whether the above bounds are compatible with the requirement
that the contribution of gravitinos to the energy density of the universe does
not exceed the observed matter density limit. From the gravitino abundance
(\ref{gabm5}) we can estimate their contribution to the closure density:
\begin{align}
\Omega_{3/2} h^2 = m_{3/2}\,\eta_{3/2}\, n_{\gamma 0}\, h^2 \rho_c^{-1}\,.
\label{closure}
\end{align}
Here $\rho_c=3 H_0^2 M_P^2/8\pi = 8.07 \times 10^{-47} h^2$~GeV$^4$ is the
critical density and $n_{\gamma 0}=3.15 \times 10^{-39}$~GeV$^3$ is the photon
density. We obtain
\begin{align}
\Omega_{3/2} h^2 \simeq 2 \times 10^{-7} \left(
\frac{M_5}{10^{10}~\mbox{GeV}}\right)^3 \left(
\frac{10^{6}~\mbox{GeV}}{T_{rh}}\right)~, \label{omegah}
\end{align}
for $m_{3/2} \simeq$ 100~GeV~-~1~TeV. Using the WMAP bound on the matter
density of the universe, $\Omega_{m} h^2 < 0.143$ \cite{Bennett:2003bz},
Eq.~(\ref{omegah}) implies the following relation between $T_{rh}$ and $M_5$:
\begin{align} \label{trhomega}
T_{rh} \gtrsim  1.4 \left( \frac{M_5}{10^{10}~\mbox{GeV}}\right)^3\,.
\end{align}
Comparing the BBN gravitino production constraint (\ref{b1t}) with the bound
(\ref{trhomega}), we see that the latter is more stringent for a gravitino mass
of the order of 1~TeV.

\section{Discussion and conclusion}

We now put together the various constraints we have derived sofar. In Figure
\ref{fig1}, we show the dependence of the 5D Planck mass $M_5$ on the effective
leptogenesis phase $\delta_L$, for two values of the gravitino mass, $m_{3/2}=
100$~GeV and 1~TeV. The lower bound on $M_5$ (solid lines) comes from the
direct leptogenesis condition, i.e. $T_{rh} < M_1$, together with
Eqs.~(\ref{mnum}) and (\ref{ntrhb}). The upper bound (dot-dashed lines) is
obtained from Eqs.~(\ref{b1t}) and (\ref{trhomega}). The shaded area is the
allowed region, which is clearly bigger for larger $m_{3/2}$. From this figure
we see that the minimum allowed values for the effective leptogenesis phase are
$\sin \delta_L \gtrsim 0.17$ and $\sin \delta_L \gtrsim 6.4 \times 10^{-3}$,
for $m_{3/2}= 100$~GeV and 1~TeV, respectively.

In Figure \ref{fig2}, we plot the reheating temperature $T_{rh}$ as a function
of $M_5$, including the bounds from gravitino production (dot-dashed) (cf.
Eqs.(\ref{b1t}) and (\ref{trhomega})), the direct leptogenesis bound (solid)
and the bound from Eq.~(\ref{trhlow}), obtained from the WMAP result for
$\eta_B$ (dotted). The shaded area corresponds to the allowed region. From
these figures we conclude that the allowed range for $M_5$ and $T_{rh}$ is
\begin{eqnarray}
3.6 &\times& 10^{10}~\mbox{GeV} \lesssim M_5 \lesssim 2.1 \times
10^{11}~\mbox{GeV}~, \nonumber\\
1.6 &\times& 10^6~\;\mbox{GeV} \lesssim T_{rh} \lesssim 9.6 \times
10^6\;~\mbox{GeV}~,
\end{eqnarray}
if $m_{3/2}= 100$~GeV, while for $m_{3/2}= 1$~TeV we find
\begin{eqnarray}
3.6 &\times& 10^{10}~\mbox{GeV} \lesssim M_5 \lesssim 5.7 \times
10^{12}~\mbox{GeV}~,\nonumber\\
1.6 &\times& 10^6\;~\mbox{GeV} \lesssim T_{rh} \lesssim 2.6 \times
10^8\;~\mbox{GeV}~.
\end{eqnarray}
Finally, the bounds on the lightest right-handed Majorana neutrino mass, $M_1$,
are the same as the ones on the reheating temperature.

Braneworld cosmology is a rich subject. During the past few years there has
been renewed activity and interest in this domain. Modifications to the
expansion rate of the universe, as is typically the case in braneworld
scenarios, can have profound implications for the processes that took place in
the early universe. In this paper, we have considered the possibility that two
of these phenomena, namely, chaotic inflation and the generation of the baryon
asymmetry of the universe through leptogenesis, occurred during the
nonconventional era in the brane. We have studied a minimal supersymmetric
seesaw scenario where the lightest singlet sneutrino field not only plays the
role of the inflaton but also produces a lepton asymmetry through its direct
decays. Taking into account the BBN constraints on the gravitino production and
the observed baryon asymmetry of the universe, we were able to strongly
constrain the fundamental 5D Planck mass scale, and consequently, the lightest
sneutrino mass, as well as the reheating temperature of the universe. The
effective leptogenesis phase is also bounded in this framework.

\begin{acknowledgments}
The work of R.G.F and N.M.C.S. was supported by  Funda\c c\~ao para a Ci\^encia
e a Tecnologia (FCT, Portugal) under the grants SFRH/BPD/1549/2000 and
SFRH/BD/4797/2001, respectively. M.C.B. acknowledges the partial support of FCT
under the grant POCTI/1999/FIS/36285.
\end{acknowledgments}

\end{document}